\documentstyle[12pt]{article}
\newcommand{\be}{\begin{equation}}
\newcommand{\ee}{\end{equation}}
\newcommand{\ba}{\begin{array}}
\newcommand{\ea}{\end{array}}
\newcommand{\ben}{\begin{enumerate}}
\newcommand{\een}{\end{enumerate}}
\newcommand{\bec}{\begin{center}}
\newcommand{\eec}{\end{center}}

\newtheorem{Thm}{Theorem}[section]
\newtheorem{Def}[Thm]{Definition}
\newtheorem{Prop}[Thm]{Proposition}

\newtheorem{Rem}[Thm]{Remark}
\newcommand{\dowod}{\noindent{\bf Proof:} }

\newcommand{\Ldwa}{\,\stackrel{2}{\bigwedge}}
\newcommand{\w}{{\!}\wedge{\!}}
\newcommand{\End}{{\rm End}\, }
\newcommand{\wg}{{\!}\wedge _g{\!}}
\newcommand{\orb}{{\cal O}}
\newcommand{\id}{{\rm id}\, }
\newcommand{\tr}{{\rm tr}\, }
\newcommand{\cx}{\overline{x}}

\font \msb=msbm10 scaled \magstep1

\newcommand{\rtimes}{\mbox{\msb o}\,}
\newcommand{\bR}{\mbox{\msb R} }
\newcommand{\bC}{\mbox{\msb C} }
\font \smsb=msbm8 scaled \magstep1
\newcommand{\smC}{\mbox{\smsb C} }

\font \eul=eufm10 scaled \magstep2

\newcommand{\gotG}{\mbox{\eul g}}

\newcommand{\gotL}{\mbox{\eul l}}

\newcommand{\X}{\mbox{\sf  X}}

\newcommand{\ar}{\alpha }

\newcommand{\dr}{\delta }

\newcommand{\lr}{\lambda }

\newcommand{\om}{\omega }
\newcommand{\vom}{\vec{\om}}
\newcommand{\vv}{\vec{v}}

\begin{document}

\title{\bf Extended phase space for a spinning particle}
\author{{\bf S. Zakrzewski}  \\
\small{Department of Mathematical Methods in Physics,
University of Warsaw} \\ \small{Ho\.{z}a 74, 00-682 Warsaw, Poland} }

\date{}
\maketitle
\begin{abstract}
Extended phase space of an elementary (relativistic) system is
introduced in the spirit of the Souriau's definition of the
`space of motions' for such system. Our `modification' consists
in taking into account not only the symmetry (Poincar\'{e})
group but also its action on the (Minkowski) space-time, i.e.
the full covariant system.  This yields a general procedure to
construct spaces in which the equations of motion can be
formulated: phase trajectories of the system are identified as
characteristics on some constraint submanifold (`mass and spin
shell') in the extended phase space.  Our formulation is
generally applicable to any homogeneous space-time (e.g. de
Sitter) and also to Poisson actions. Calculations concerning the
Minkowski case for non-zero spin particles show an intriguing
alternative: we should either accept two-dimensional
trajectories or (Poisson) noncommuting space-time coordinates.

\end{abstract}

\section{Introduction}

According to Souriau \cite{Sou}, the space of `motions'
(`histories', `phase trajectories') of a classical mechanical
system has a structure of a symplectic manifold. If the system
is isolated, then the space-time symmetry group acts
(symplectically) on this manifold. {\em Elementary systems} are
those for which this action is transitive (such systems `do not
have other structure than their space-time situation'
\cite{Sou}).  By the momentum mapping theory, transitive actions
correspond to coadjoint orbits of the underlying group (modulo
possible cohomological problems, not present in our basic case:
Poincar\'{e} group).

This is probably the most basic physical application of groups.
The symmetry group under question (Poincar\'{e}, Galileo, de
Sitter,...) determines (by an algorithm) possible types of
elementary particles (mass, spin,...) and the set of their
motions.  However, the motions are described only as abstract
points of coadjoint orbits. The algorithm does not provide any
description of motions as solutions of `equations of motion',
which we'd expect to be formulated in an appropriate bundle over
space-time.

As a way to determine the full model of particle, we introduce
in this paper {\em extended phase spaces} (in which the
equations of motion can be formulated). Like the `space of
motions', an extended phase space is defined as a symplectic
`transitive' space, the transitivity this time being understood
with respect to the pair {\em group}~+~{\em space-time} rather
than to the group alone (we `represent' not only the
infinitesimal generators of the group but also functions on the
space-time, in a covariant way).
%  also: to deal directly with space-time localization

The paper is organized as follows.  The definition of extended
phase spaces is given in Section~\ref{cov}, after a short
investigation of the simplest well known case in
Section~\ref{homo}. All extended phase spaces are classified in
Section~\ref{class} (they turn out to be in one-to-one
correspondence with the coadjoint orbits of the Lorentz group).
The reduction of an extended phase space by fixing value of {\em
spin} and {\em mass} (which relates the extended phase space to
a coadjoint orbit of the Poincar\'{e} group) is described in
Section~4. It turns out that the history of the particle is
represented by a `world tube' rather than world line.

It is convenient to have in mind here also the quantum case.
Quantum elementary particles are related to irreducible unitary
representations of the space-time symmetry group and the
corresponding full description (wave equation) should be (in our
approach) related to covariant representations of the pair {\em
group}~+~{\em space-time} (in this context, such a pair is
called a {\em dynamical system}; we prefer to avoid this
terminology). In Section~5 we explain how the quantum formulation
results from the classical one.

In our opinion, it is convenient to use similar language both in
the quantum and the classical case. We shall speak about
representations, irreducibility, etc. in the classical case
(instead of symplectic realizations, transitivity, etc.).

In Section~6 we perform a partial reduction of an extended phase
space by fixing the value of spin. The original
fibration over space-time does not descend to the quotient, but
there exist another one which does it. With respect to the new
fibration, the space-time coordinates do not commute, the
Poisson bracket being proportional to the spin tensor and
inverse proportional to the square of mass.

\section{Homogeneous formulation of mechanics and relativistic
spin zero particle}\label{homo}

Let $Q$ denote the configuration space of a non-relativistic
mechanical system. A (non-homogeneous) {\em hamiltonian formulation
of dynamics} of such a system is given by specifying a
(time-dependent) Hamiltonian function $H\colon \bR \times T^* Q\to
\bR $, which generates the equations of motion:
\be\label{ham}
\dot{x}^k = \frac{\partial H}{\partial p_k},\qquad \dot{p}_k =
-\frac{\partial H}{\partial x^k}.
\ee
Here $T^* Q$ denotes the cotangent bundle of $Q$ (the phase space),
 $(x^k)_{k=1,\ldots ,N}$ -- some
coordinates in $Q$, $(x^k,p_j)$ -- the induced coordinates on
the phase space and the dot denotes the differentiating with respect
to time.

Now consider the {\em extended configuration space}
${\cal M}:=\bR \times Q$ and the hypersurface
$C_H$ in $T^* {\cal M}$  given by
\be
         C_H = \{ (t,x^1,\ldots ,x^N,e,p_1,\ldots ,p_N)\in T^*
{\cal M} :
e = H(t,x,p)\}
\ee
(`energy' = `Hamiltonian'), where $t=x^0$ is the time variable and
$-e=p_0$ is the conjugate variable (minus `energy'). Of course,
specifying $H$ is the same as specifying $C_H$. It is easy to check
that solutions of (\ref{ham}) are in one-to-one correspondence with
characteristics on $C_H$ (characteristics = the integral curves
of the degeneracy distribution of the symplectic form restricted
to $C_H$). The description in terms of $C_H$ is said
to be the {\em homogeneous formulation of hamiltonian dynamics} (cf.
e.g. \cite{BT}). The cotangent bundle $T^* {\cal M}$ is said to
be the {\em extended phase space}.

The homogeneous description is particularly useful in the case of a
relativistic point particle (with spin zero). In this case
${\cal M}$
is just the Minkowski
space-time $M$:
$$ {\cal M} := M.$$
 For a free particle with mass $m$, the corresponding
submanifold of $T^* M$ is just the `mass shell':
\be\label{ms}
      C_m = \{ (x^0,x^1,x^2,x^3,p_0,p_1,p_2,p_3): p^2 = m^2
,\; p_0>0 \}.
\ee
Here $p^2 = g^{kl}p_{k}p_{l}$ is the Lorentz square of the
4-momentum ($g^{kl}$ is the contravariant Lorentz metric).

The Poincar\'{e} group --- the (connected) group of affine
transformations of $M$ leaving $g$ invariant --- will be denoted
by $G$, its Lie algebra --- by $\gotG $. The canonical moment
map $J\colon T^* M\to \gotG ^*$ for the action of $G$ on
$T^*M$ (the lift of the natural action of $G$ on $M$) identifies
the set of characteristics on $C_m$ with the coadjoint orbit in
$\gotG ^*$ corresponding to the mass $m$ and spin zero.
This gives a natural realization of the abstract points of this
coadjoint orbit as trajectories. The equations of motion are
encoded in the mass shell $C_m$ (which is nothing else but the
inverse image of the coadjoint orbit by $J$).

We regard the above description as a full model of a (free)
relativistic particle with mass $m$ and spin zero. Now we
extract its main features in order to pass to a general case.
We observe the following three essential properties of the above
model (we set $P:=T^*M$):
\ben
 \item $P$ is a Hamiltonian $G$-space, in other words,

 \vspace{2mm}

 \bec \fbox{a complete Poisson map $J\colon P\to \gotG ^*$ is
given} \eec

 \item $P$ is fibered over $M$ (with coisotropic fibers), i.e.

 \vspace{2mm}

 \bec \fbox{a complete Poisson map $\pi\colon P\to M$ is
given} \eec

 \item the following covariance holds: \ \ $X_P ( \pi ^* f )=\pi
^* (X_M f)$, or, equivalently,
\be\label{cova}
\mbox{\fbox{$\{ J^* X, \pi ^* f\} = \pi ^* (X_M f)$} }
\ee
for $X\in \gotG$, $f\in C^{\infty}(M)$. Here $X_M$ (or $X_P$)
denotes the fundamental vector field of the action of $G$ on $M$
(or $P$), corresponding to $X\in \gotG$, and $\pi ^* f$ is the
pullback of $f$ by $\pi$ (similarly, $J^*X$ is the pullback of
$X$ by $J$, where $X$ is treated as a linear function on $\gotG
^*$). Of course for $P=T^*M$, $\pi $ is the cotangent bundle
projection.

\een
\begin{Rem}
{\rm We recall that $\gotG^*$ is naturally a Poisson manifold. The
Poisson structure on $M$ is zero. A Poisson map is said to be}
 complete {\rm \cite{qcp}, if it sends (by pullback) functions
having complete Hamiltonian vector fields on functions with the
same property (such functions are called complete).}
\end{Rem}

\section{Covariant representations, extended phase spaces}
\label{cov}

It is convenient to introduce the following terminology.
\begin{Def}
{\rm A} representation of a Poisson manifold $N$ in a symplectic
manifold $P$ {\rm is a complete Poisson map $\Psi $ from $P$ to $N$.}
 \end{Def}
 \begin{Def}
{\rm Suppose we are given an action of a Lie group $G$ on a
manifold $M$. A } covariant representation of $(M,\gotG ^* )$ in
a symplectic manifold $P$
{\rm is a pair $(\pi ,J)$, where $\pi $ is a representation of $M$
in $P$, $J$ is a representation of $\gotG ^*$ in $P$ and the
 condition of covariance (\ref{cova}) is satisfied.}
\end{Def}
 An example of a covariant representation was presented in the
previous section. In fact, it has one more important property:
it cannot be `decomposed' onto smaller `subrepresentations', because
\be\label{irre}
\mbox{ \fbox{$X_P$, $\X _{\pi ^*f} $ (with $X\in \gotG$,
$f\in M$) span $TP$}}.
\ee
Here $\X _{h}$ denotes the Hamiltonian vector field of the
function $h$.
Note that $X_P=\X _{J^*X}$ for $X\in \gotG\simeq (\gotG ^*)^*$
and one can replace $X_P$ in (\ref{irre}) by $\X _{J^*\phi }$
for $\phi\in C^{\infty} (\gotG ^*)$.

 We say that a covariant representation $(\pi ,J)$
of $(M,\gotG ^*  )$ in $P$ is {\em irreducible} if condition
\ref{irre} is satisfied. Similarly,
a representation $\Psi $ of a Poisson manifold $N$ in a symplectic
manifold $P$ is said to be {\em irreducible}, if
$\X _{\Psi ^* h}$ span $TP$ for $h\in C^{\infty}(N)$.

We can now introduce our fundamental definition.
\begin{Def}
{\rm By an} extended phase space of a relativistic particle {\rm
we mean an irreducible covariant representation of $(M, \gotG
^*)$, where $M$ is the Minkowski space and $G$ is the
Poincar\'{e} group.}
\end{Def}
\begin{Rem}
{\rm The above definition is applicable to other situations,
like the de Sitter space-time or the case of Poisson Minkowski
space \cite{poihom}. In the latter case one should replace
$\gotG ^*$ by $G^*$ --- the Poisson dual of the Poisson
Poincar\'{e} group \cite{poi,poican,PPgr,k-part}, and also $J^*X$
in (\ref{cova}) --- by the right-invariant 1-form on $G^*$
corresponding to $X$.
 In
all these cases one
has the basic example provided by the cotangent bundle of $M$
(symplectic groupoid \cite{We:sg,CDW:gs,Ka,qcp} of $M$ in the
case of general Poisson $M$, see
\cite{poi,poican,k-part,abel}).}
\end{Rem}

 In order to find all covariant representations of $(M,\gotG
^*)$ for a given action of $G$ on $M$, we notice that they are
in 1--1 correspondence with representations of a certain Poisson
manifold (similar fact is known in the theory of crossed
products).
\begin{Prop} There is a 1--1 correspondence between covariant
representations $(\pi ,J)$ of $(M,\gotG ^*)$ and representations
$\Psi $ of the semi-direct Poisson product $M\rtimes\gotG ^*$,
given by
$$  \Psi = \pi \times  J.$$
$\Psi $ is irreducible if and only if $(\pi ,J)$ is irreducible.
\end{Prop}
To convince that this proposition is reasonable,
recall \cite{We:semi} that $M\rtimes \gotG ^*$ is the cartesian
product of $M$ and $\gotG ^*$ equipped with the {\em semidirect
Poisson structure} defined by
\be\label{semi}
\{ f_1,f_2\} =0,\qquad \{ X_1,X_2\} =[X_1,X_2],\qquad \{ X,f\}
=X_Mf
\ee
for $f_1,f_2\in C^{\infty} (M)$, $X_1,X_2\in \gotG $. We choose
the convention that $X\mapsto X_M$ is a homomorphism of Lie
algebras, hence we choose the commutator in
$\gotG$ based on {\bf right-invariant vector fields}.

Since irreducible representations of a Poisson manifold are just
(coverings of) its symplectic leaves (this is a generalization
of the familiar fact concerning the moment map of a transitive
hamiltonian action), we conclude that irreducible covariant
representations of $(M,\gotG ^*)$ are (coverings of)
symplectic leaves in $M\rtimes \gotG ^*$.

\section{The classification of extended spaces} \label{class}

In order to describe all possible extended phase spaces we have
to study the structure of the Poisson manifold $M\rtimes \gotG^
*$ in the Minkowski-Poincar\'{e} case.

We denote by $V$ the subgroup of translations in the
Poincar\'{e} group $G$. This is a normal subgroup and
$$ L:=G/V$$
is the Lorentz group, acting naturally in $V$ --- the tangent
space of $M$. Any choice of $x\in M$ allows to identify $L$ with
the stabilizing subgroup $G_x$ of $G$. We denote by $\gotL $ and
$\gotG _x$ the
 Lie algebras corresponding to $L$ and $G_x$.
\begin{Prop}
The natural map
$$ M\rtimes \gotG ^* \ni (x,\ar )\mapsto ((x,p), S)\in
T^*M\times \gotL ^*,$$
where $p$ is the restriction of $\ar $ to $V$ and $S$ is the
restriction of $\ar$ to $\gotG _x \simeq \gotL $, is a Poisson
isomorphism ($T^*M\times \gotL ^*$ considered with its direct
product Poisson structure).
\end{Prop}

\dowod Choose a basis $e_k$ in $V$ and set
$$M_{kl}:=e_k\otimes g(e_l) -e_{l}\otimes g(e_k)\in \gotL\subset
\End V.$$
The `right' commutators in $\gotG \simeq V\rtimes \gotL$ (we fix
an identification $M\simeq V$) are given by
$$ [M_{jk},M_{ln}] = M_{jl}g_{kn} +  M_{kn}g_{jl} -
M_{jn}g_{kl} - M_{kl}g_{jn},\qquad [M_{jk},e_{l}] = -e_jg_{kl} +
e_kg_{jl}.$$
The same formulas define the Poisson brackets on $\gotG ^*$:
$$ \{ M_{jk},M_{ln}\} = M_{jl}g_{kn} +  M_{kn}g_{jl} -
M_{jn}g_{kl} - M_{kl}g_{jn},\qquad \{ M_{jk},p_{l}\} =
-p_jg_{kl} + p_kg_{jl} $$
 (the elements of $\gotG$ are now (linear) functions on $\gotG
^*\simeq V^*\times \gotL ^*$).
We have denoted $e_k$, viewed as functions on $V^*$, `more
physically' --- by $p_k$ (the momenta).

It is easy to see that the `cross' Poisson brackets in $M\rtimes
\gotG ^*$ are given by
$$\{ M_{jk},x^l\}= - x_j\dr _k{^l} + x_k\dr _j{^l},\qquad \{
p_j, x^l \} = \dr _j{^l} ,$$
where $x^l$ are coordinates on $M\simeq V$ (corresponding to
$e_l$) and $x_k = g_{kl}x^l$ (summation convention).

The transformation $(x,(p,M))\mapsto ((x,p),S)$ is given in
terms of coordinates by
$$ S_{jk} = M_{jk} - p_jx_k + p_kx_j .$$
Now it is easy to see that $\{ S_{jk},x^l\} = 0$, $\{
S_{jk},p_l\} = 0$ and
\be\label{Snaw}
 \{ S_{jk},S_{ln}\} = S_{jl}g_{kn} +  S_{kn}g_{jl} -
S_{jn}g_{kl} - S_{kl}g_{jn} .
\ee
The latter equality follows easily from
$$ \{ w_{jk},w_{ln}\} = w_{jl}g_{kn} +  w_{kn}g_{jl} -
w_{jn}g_{kl} - w_{kl}g_{jn}, $$
where $w_{jk}:= p_jx_k - p_k x_j$ (a consequence of the fact
that $w_{jk}$ describes the canonical momentum mapping for the
action of $L$ on $T^*M = T^*V$), and
$$ \{ M_{jk},w_{ln}\} = w_{jl}g_{kn} +  w_{kn}g_{jl} -
w_{jn}g_{kl} - w_{kl}g_{jn} .$$

\hfill $\Box $

\noindent
{\bf Corollary.} \ Extended phase spaces are of the form
$$  P = T^*M \times \orb ,$$
where $\orb $ is a coadjoint orbit in $\gotL ^*$. They are in
one-to-one correspondence with these coadjoint orbits. The
trivial coadjoint orbit yields simply $T^*M$ --- the extended
phase space of a spinless particle,
described in Sect.~\ref{homo}.

\vspace{1mm}

Now we recall some basic facts concerning the Lorentz Lie
algebra $\gotL$. By definition, $\gotL\subset \End V$ is the
orthogonal Lie algebra of the Lorentz metric $g$ of signature
$(1,3)$ in $V$. The map
$$\id\otimes g : V\otimes V\to V\times V^*\equiv \End V$$
defines a linear isomorphism between $\Ldwa V$ and $\gotL$. We
set
$$  x\wg y := x\otimes g(y) - y\otimes g(x) = (\id\otimes g)(x\w
y)$$
for $x,y\in V$.

We shall identify $\gotL$ with its dual --- our `spin
variable' $S$ will then take values in $\gotL$ --- using the
invariant form
$$ < S,S>:= \frac12 \tr S^2 .$$
Any `timelike' vector $u\in V$ such that $g(u,u)=1$ defines the
orthogonal decomposition of $\gotL$ on `rotations' and `boosts':
$$ \gotL = \gotL _u + (\gotL _u)^{\perp} ,$$
$$ S = (S - Su\wg u) + Su\wg u \qquad \mbox{for $S\in \gotL$}.$$
The boost part, $Su\wg u$, is encoded in the vector $v:= Su$
belonging to $u^{\perp}$ --- the orthogonal complement of $u$.
The rotation part, $S - Su\wg u$, is also encoded in a vector
$\om\in u^{\perp}$, such that
$$ (S - Su\wg u) x = \om \times x\qquad \mbox{for}\;\; x\in
u^{\perp} ,$$
where $\times$ denotes the three-dimensional vector product in
$u^{\perp}$ (suppose we fix an overall orientation). Therefore,
we can represent $S\in \gotL$ by a pair of vectors, $(\om ,v)$,
where $\om ,v\in u^{\perp}$. Using the bijection
between $\gotL _u$ and $(\gotL _u )^{\perp}$ (both are isomorphic to
$u^{\perp}$), one can see that $\gotL \simeq (\gotL _u)^{\smC}$.
The appropriate complex structure on $\gotG$ is given by the
following `multiplication by $i$':
$$ J(\om ,v) := ( -v,\om ).$$
We conclude that $\om$ is calculated from $S$ as follows: $\om =
(JS)u$ (recall that $v=Su$).

In terms of components $\om $, $v$, the Killing form reads
$$ <S,S> = g(\om ,\om ) - g(v,v)= -(\vom ^2 -\vv ^2 ) ,$$
where $\vom ^2:= - g(\om ,\om )$ denotes the positive definite
metric in the three-dimensional space.
We have
$$ < JS,S>= -2 g(\om ,v)= 2\vom \cdot \vv ,$$
and the {\em complex} invariant form on $\gotL$  which extends
the previous real form on $\gotL _u$ is given by
$$<S,S>_{\smC} = <S,S> -i<JS,S> = -(\vom ^2 -\vv ^2) - 2i
\vom\cdot \vv = - (\vom + i\vv )^2 .$$
Knowing that $\gotL\simeq  sl(2,\bC )$, one can easily see that
the value of the complex Killing fully specifies the adjoint
orbit (if we consider only nontrivial orbits).
Therefore, for any complex number $z= a+ib$ we have the orbit
$$\orb _z := \{ S\in \gotL \setminus \{ 0\} : <S,S>_{\smC} = -
z^2\}= \left\{ (\om , v) : \ba{rcl} \vom ^2 - \vv ^2 & = & a^2 -
b^2 \\ \vom \cdot \vv & = & ab \ea\right\} $$
($ z$ and $-z$ correspond to the same orbit). Each such
orbit is of dimension 4. Corresponding extended phase spaces are
12-dimensional.

\section{Fixing spin and mass}

Each extended phase space $P_z=T^*M\times \orb _z$ decomposes onto
orbits of the
Poincar\'{e} group. Generically, they are obtained by fixing
values of the two invariants: spin and mass. We shall discuss
only the case of the positive value of the mass square:
$$ m^2:=g(p,p)>0.$$
For simplicity, we identify the momentum $p\in V^*$ with the
corresponding vector $g^{-1}(p)\in V$. The spin $s$ is given by the
following expression
$$ s^2 = -g((JS)u,(JS)u)= \vom ^2 ,$$
where $u:= \frac{p}{m}$ is the unit vector in the direction of
$p$ (recall that the Pauli-Lubanski vector is defined by
$W=(JS)p$).

Fixing $m$ and $s$, we obtain generically a 10-dimensional
coisotropic submanifold (`spin and
mass shell') in $P_z$. The characteristic foliation on this
submanifold is therefore 2-dimensional. In order to find the
leaves of this foliation, it is sufficient to integrate the
hamiltonian vector fields of $\frac12 m^2$ and
$\frac12 s^2$. Since
$$\{ \frac12 m^2 ,x\} =p,\qquad \{\frac12 m^2,p\} =0, \qquad
\{\frac12 m^2 ,S\}=0,$$
the mass constraint generates the usual rectilinear motion with
conserved $p$ and $S$ and four-velocity $u$.
In order to calculate the Poisson brackets with $\frac12 s^2$,
note, that
$$ s^2 = g(S^2 u, u) + <S,S>.$$
A simple calculation yields then
$$\{\frac12 s^2 ,p\} =0,\qquad \{\frac12 s^2 ,x\}
=\frac{1}{m}(S^2 u\wg u) u,\qquad \{\frac12 s^2 ,S\}
= -S^2 u\wg u\in (\gotL _u)^{\perp}$$
(in order to compute the last Poisson brackets, note, that the
matrix elements $S^j{_k}$ of $S$, which are functions on
$\gotL$, correspond via the chosen invariant form  to
$-g^{jl}e_l\wg e_k$ hence their
Poisson brackets are minus the standard ones (\ref{Snaw})).
It follows that $p$ and $\om$ (the rotational part of $S$) are
conserved. Since $S^2 u =Sv = \vom\times \vv + \lr u$, we have
$S^2 u\wg u = (\vom\times\vv )\wg u$, hence the flow of $\frac12
s^2$ changes $v$ (the boost part of $S$) according to
$$ \{\frac12 s^2 ,\vv \} = \vom\times\vv .$$
It means that $\vv$ simply rotates around the $\vom$ axis.
Since
$$\{\frac12 s^2 ,x+\frac1{m}Su\}=0,$$
 vector
$$ \cx := x +\frac1{m} Su ,$$
is conserved, and $x$ moves on a circle around the axis passing
through $\cx$ in the direction of $\vom$:
$$ x = \cx -\frac1{m} Su = \cx - \frac1{m} v .$$
We conclude that the characteristics have the form of a
2-dimensional cylinder. Their projections on the Minkowski
space-time are then `world tubes' rather than `world lines'.
Each of these world tubes corresponds to a point of the
coadjoint orbit in $\gotG ^*$ (with the fixed value of $m$ and
$s$) and should represent the `history' of the elementary
system. In this sense we have obtained `two-dimensional
trajectories'.
In a co-moving frame, an observer
should see a circle of the radius $r$ with
$$ r^2 = \frac{(\vv ^{\perp})^2}{m^2} = \frac{1}{m^2} \left( \vv
^2 - \frac{(\vom \cdot \vv )^2}{\vom ^2}\right) =
\frac{1}{m^2s^2}\left( s^2(s^2 - a^2 + b^2) - a^2b^2\right) $$
($\vv ^{\perp}$ denotes the component of $\vv$ perpendicular to
$\vom$), i.e.
$$ r^2 = \frac{1}{m^2s^2}(s^2 - a^2)(s^2 + b^2).$$
For a fixed radius $r$ and orbit $\orb _z$, $z=a+bi$, the above equation
imposes a relation between spin and mass, asymptotically linear
 ({\em Regge trajectory?}).

 \section{Quantization}

 The scheme we have presented is sufficiently universal in order
to describe immediately the quantum case. Let us consider for
example the case $z=0$. One can show that
$$\orb _0 \simeq T^*{\cal S}^2 \setminus {\cal S}^2$$
--- the cotangent bundle to the 2-sphere without the image of
the zero section, where the cotangent bundle polarization
corresponds to an invariant polarization on $\orb _0$. The sphere
${\cal S}^2$ here is in fact the {\em celestial sphere}
(projective forward light cone in $V$).
Neglecting the measure zero set we have
$$P_0 = T^* M \times \orb _0 \simeq T^* (M\times {\cal S}^2),$$
whose quantum counterpart is
$$ L^2 (M)\otimes L^2 ({\cal S}^2)= L^2 (M\times {\cal S}^2).$$
Now the wave equations corresponding to the `mass and spin
shell' are simply obtained by replacing the classical quantities
by the quantum ones:
\be\label{quant}
\left\{ \ba{rcl} \Box _x \psi (x,\theta ) & = & m^2 \psi
(x,\theta ) \\ \hat{W} ^2 \psi (x,\theta ) & = & m^2 s(s+1)\psi
(x,\theta )\ea \right.
\ee
Here $\Box _x$ is the d'Alembert operator with respect to the
$x$ variable, $\theta $ denotes the variable on ${\cal S}^2$ and
$\hat{W} ^2$ arises from
$$ W^2 = - g(W,W) = - g((JS)p,(JS)p) = g((JS)^2p,p) = g(S^2p,p)
+ <S,S>g(p,p) $$
$$= -g_{jk}S^j{_l}p^lS^k{_n}p^n + <S,S>m^2$$
by replacing $p_{k}$ by $-i\frac{\partial}{\partial x^k}$ and
$S^j{_l}$ by the generators of the representation of the Lorentz
group $L$ in $L^2({\cal S}^2)$.

The simplest orbit $\orb _0$ corresponds to a unitary
representation of the Lorentz group which contains only integral
spins (the minimal spin is zero). In order to be able to pick up
also half-integral spins, one has to consider other
orbits/representations. Orbits $\orb _z$ with $z\neq 0$, are
known to be lagrangian bundles over ${\cal S}^2$ (affine bundles
modelled on $T^*{\cal S}^2$). Using geometric quantization, one
can construct (for quantizable orbits, i.e. for $a$ being
half-integer) the corresponding
unitary representations of the Lorentz group (or, rather its
universal cover, $SL(2,\bC )$) and then wave functions on ${\cal
S}^2$ have to be replaced by sections of a suitable complex line
bundle over ${\cal S}^2$. This way one obtains the principal
series of representations of $SL(2,\bC )$ numbered by two
parameters:  one discrete and one continuous.
The orbit $\orb _z$, $z=a+bi$, corresponds to the unitary
representation of $SL(2,\bC )$ induced from the representation
of the parabolic subgroup:
$$\left(\ba{cc} \lr & 0 \\ \mu & \lr ^{-1}\ea\right)\;
\mapsto \; |\lr |^{2ib}\left(\frac{\lr}{|\lr |}\right)^{-2a}=\lr
^{-2a} (\lr\overline{\lr})^{a+bi} $$
($a$ is half-integer).

{\bf Remark.} \
In order to obtain the quantum case, it is not necessary to
consider first the classical case and then to worry about
correct quantization. We can just consider any irreducible
unitary representation of $L$  in a Hilbert space $H$, and pick
up (by any means) a (generalized) irreducible subrepresentation
of $G$ in $L^2(M)\otimes H$, which, essentially, amounts again
to equations (\ref{quant}).

{\bf Problem:} \ What is the direct relation between solutions
of (\ref{quant}) and solutions of wave equations in some
standard formulation?

\section{Fixing spin only}

It is natural to look for a possibility to fix spin first in
order to obtain a 10-dimensional reduced symplectic manifold.
In this manifold we could then consider pure mass shell
(being more close to the concept of a wave equation in the
traditional sense).

We thus consider the submanifold
$${\cal C}_{z,s} =\{ {\rm spin} = s \} \subset (P_z)_+ = (T^*
M)_+\times
\orb _z \subset P_z$$
of a fixed spin. Here $(T^*M)_+ = M\times V^*_+$ is the subset
of $T^*M$ corresponding to time-like momenta.

We recall that the characteristics on ${\cal C}_{z,s}$ are
topological circles whose projection on $M$ are
circles (in the co-moving frame) of the radius $r$ given by
$$ r^2 = \frac{1}{m^2s^2}(s^2 - a^2)(s^2 + b^2).$$
It follows that the spin function is bounded from below on $(P_z)_+$:
$$ s\geq |a|.$$
We have then two cases.
\ben
\item $s>|a|$. In this case $r>0$, hence $\X _{s^2}\neq 0$
(characteristics really exist), $d (s^2)\neq 0$ and
 $\dim {\cal C}_{z,s} = 11$ (${\cal C}_{z,s}$ is
coisotropic). The projections of characteristics on $M$ are
`circles' (not points), therefore the variables $x^k$ {\bf do
not pass} to the quotient
$$ P_{z,s}:= {\cal C}_{z,s}/\{ {\rm circles}\}.$$
Still, the `renormalized' position
$\cx = x -\frac1{m} Su$ is of course well defined on $P_{z,s}$.
Since we have coisotropic constraints, the Poisson bracket of
$\cx ^j$ and $\cx ^k$ in $P_{z,s}$ is equal to their Poisson
bracket in $P_z$. The calculation gives
\be\label{spinkomu}
\{\cx ^j,\cx ^k\} = \frac{1}{m^2}(S^{jk} - (Su)^ju^k + (Su)^k
u^j)= \frac{1}{m^2}R^{jk} ,\qquad (m^2\equiv p^2)
\ee
where $R:= S - Su\wg u$ is the rotation part of $S$ (with
respect to $u$). The full description of $P_{z,s}$ can be given
in terms of $\cx ^j$, $p_k$, $R\in \gotL$ (such that $Ru=0$,
$<R,R>= -s^2$) and Poisson brackets
$$\{p_k,\cx ^j\}=\dr _k^j,\qquad \{p_k,p_j\}=0,\qquad
\{\cx ^j,\cx ^k\} = \frac{1}{m^2}R^{jk} ,\qquad \{
p_k,R_{jn}\}=0,$$
$$ \{R_{jk},\cx ^l\} = \frac{1}{m}(R_{kl}u_j-R_{jl}u_k),
\qquad
\{ R_{jk},R_{ln}\} = -(R_{jl}\tilde{g}_{kn} +
R_{kn}\tilde{g}_{jl} - R_{jn}\tilde{g}_{kl} -
R_{kl}\tilde{g}_{jn}) ,$$
where $\tilde{g} _{jk}=g_{jk}- u_ju_k $ is the three-dimensional
metric.
\item $s=|a|$. In this case $\vom ^2 = a^2$, hence $\vv ^2 =
b^2$. Since $\vom\cdot\vv = ab$, $\vv \parallel \vom$ and it is
easy to see that
 $\dim {\cal C}_{z,s} = 10$. Calculating the symplectic form in
$P_{z,s}$ on vectors tangent to  ${\cal C}_{z,s} $ we obtain the
following results
 \ben
  \item for $a\neq 0$,  ${\cal C}_{z,s}$ is a symplectic
(sub)manifold ({\em second class} constraints!) and
\be\label{spinkomu2}
\{ x ^j,x ^k\} = \frac{1}{m^2}(1+\frac{b^2}{a^2})R^{jk}.
\ee
In particular, when $b=0$, $\{ x^j,x ^k\} = \frac{1}{m^2}R^{jk}
= \frac{1}{m^2}S^{jk}$ ($R=S$ in this case).
   \item  for $a = 0$,  ${\cal C}_{z,s}$ is coisotropic and
    $P_{z,s} \simeq T^*M$ (the spinless case).
  \een

\een

\section{Conclusions}

We have constructed extended phase spaces as symplectic
manifolds endowed with a Hamiltonian action of the Poincar\'{e}
group and carrying a localization structure. Trajectories of an
elementary system are characteristics on the `spin and mass'
shell in the extended phase space. They are typically
2-dimensional, due to the fact that we impose two constraints.
The value of spin is related to the radius $r$ of the
world tube. For big values of $s$, or for $P_0$, this relation
looks as follows
$$ s = mr .$$
This reminds the orbital angular momentum of a particle with
(effective, not rest) mass $m$ moving on a circle of radius $r$
with the velocity of light ($c=1$).

An attempt to introduce an `intermediary extended phase space'
(by a reduction with respect to a fixed spin) which would be
still fibered over space-time (to this end we have to modify the
original fibration), leads to `non-commutative' space-time. The
non-commutativity holds between the coordinates in the
two-dimensional subspace orthogonal to the four-velocity and the
actual direction of spin (the subspace of rotation). The proper
angular momentum plays therefore the role of the `source of
non-commutativity'.

In the case of the extreme value of spin on $P_z$, the original
fibration over space-time does not have to be modified. However,
since the reduction is not coisotropic in this case, the
original positions no longer commute (as functions on the
constraint manifold). The commutation rules have here the form
similar to the previous ones, with spin being the source of the
non-commutativity. Since the reduction is not coisotropic, we do
not have, unlike before, the `mechanical' explanation of the
non-commutativity in terms of replacing the commuting control
parameters (canonical positions) by new control parameters
(kinetic positions), no longer commuting, chosen for the reason
of good transformation properties (in this connection, see also
\cite{poican}).

\end{document}